\documentclass[preprint,superscriptaddress,noshowpacs,longbibliography]{revtex4-1}
\usepackage{amssymb}
\usepackage{amsmath}
\usepackage{amsfonts}
\usepackage{bm}
\usepackage{graphicx}
\usepackage[dvipsnames]{xcolor}
\usepackage{calc} % to do for example 0.5\columnwidth+0.2\columnsep
\usepackage{epsfig}
\usepackage{epstopdf}
\usepackage{natbib}
\usepackage{hyperref}
\begin{document}

\title{Realizing universal quantum gates with topological bases in quantum-simulated  superconducting chains}

\author{Yong Hu}
\email{huyong@mail.hust.edu.cn}
%\email{huyong@hku.hk}
\affiliation{School of Physics, Huazhong University of Science and Technology,
Wuhan, 430074, China}
\affiliation{Department of Physics and Center of Theoretical and Computational
Physics, The University of Hong Kong, Pokfulam Road, Hong Kong, China}

\author{Yu-Xin Zhao}
\email{yuxinphy@hku.hk}
\affiliation{Department of Physics and Center of Theoretical and Computational
Physics, The University of Hong Kong, Pokfulam Road, Hong Kong, China}

\author{Zheng-Yuan Xue}
\email{zyxue@scnu.edu.cn}
\affiliation{Guangdong Provincial Key Laboratory of Quantum Engineering and Quantum Materials, School of Physics and Telecommunication Engineering, South China Normal University, Guangzhou 510006, China}
\affiliation{Department of Physics and Center of Theoretical and Computational
Physics, The University of Hong Kong, Pokfulam Road, Hong Kong, China}

\author{Zi-Dan Wang}
\email{zwang@hku.hk}
\affiliation{Department of Physics and Center of Theoretical and Computational
Physics, The University of Hong Kong, Pokfulam Road, Hong Kong, China}

\begin{abstract}
One-dimensional time-reversal invariant topological superconducting wires of the symmetry class DIII exhibit exotic physics which can be exploited to realize the set of universal operations in topological quantum computing. However, the verification of DIII-class physics in conventional condensed matter materials is highly nontrivial due to realistic constraints. Here we propose a symmetry-protected hard-core boson simulator of the one-dimensional DIII topological superconductor. By using
the developed dispersive dynamic modulation approach, not only the
faithful simulation of this new type of spinful superconducting chains is achieved, but also a set of universal quantum gates can be realized with the computational basis formed by the degenerate ground states that are topologically protected  against random local perturbations. Physical implementation of our scheme based on a Josephson quantum
circuit is presented, where our detailed analysis pinpoints that this scheme is experimentally feasible with the state-of-the-art technology.
\end{abstract}

%\pacs{03.67.Lx, 03.67.Ac, 71.10.Fd, 85.25.-j }
\maketitle

\section*{Introduction}
\label{Sec Intro}
Topological band theory has become the focus of recent research due to its significant importance in fundamental physics and potential applications in novel devices \cite{HasanRMP2010,QiZhangRMP2011,BernevigBook}. In each spatial dimension, topological insulators and topological superconductors can well be classified in terms of the presence or absence of time-reversal symmetry (TRS) and particle-hole symmetry \cite{Schnyder-Classification,Kitaev-Classification}. Typical examples include the integer quantum Hall effect \cite{KlitzingPRL1980} belonging to the A class of 2D and the Kitaev chain belonging to the D class of 1D \cite{Kitaev}. Among the various symmetry-protected topological
phases, increasing research interests have been drawn to the theoretical and
experimental studies of 1D topological superconducting wires in the symmetry
class DIII~\cite{DIII-I,DIII-II,DIII-III,DIII-IV,ZhaoWang-MF,DIII-V,DIII-VI}. In contrast to the Kitaev chain in the class D with one Majorana zero mode
(MZM) at each end and two-fold degenerate ground states in its $\mathbb{Z}%
_{2}$ topological phase \cite{Kitaev}, the DIII superconductor wire in
the $\mathbb{Z}_{2}^{(2)}$ topological phase possesses the TRS, and has a Kramers doublet of MZMs at each end and four-fold
degenerate ground states~\cite{ZhaoWang2013,ZhaoWang2014,Zhao-Wang-Z2}. Moreover, the DIII superconductor
wire exhibits  an impactful response to an effective magnetic field and long-range
spin-correlation at the two ends with the fixed fermionic parity, which can be
utilized to realize universal quantum operations of qubits that are topologically protected against random local perturbations
\cite{ZhaoWang-MF,DIII-V,DIII-II}. However, despite the extensive theoretical efforts, still the 1D TRS-associated $\mathbb{Z}_{2}^{(2)}$
topological phase has not been tested experimentally in electronic systems as it is limited by realistic reasons, e.g. the constraints of materials, the lack of controllability, and the co-existing complicated mechanisms.

On the other hand, it has been indicated that a kind of  one-dimensional (1D)
hard-core boson (HCB) chain may be able to simulate fermionic physics
\cite{FermionizedPhoton}, with interesting phenomena being investigated
including the pursuit of Majorana fermions in the Kitaev model \cite%
{ImamogluMajorana,WangMajorana,MaoMajorana}, the string breaking dynamics of
quark pairs \cite{ZollerGauge}, the electron-electron scattering \cite{FermionScatter}, and the delocalization of Dirac fermions in disordered one-dimensional wires \cite{SLZhu2009}. However, methods in the existing studies are restricted to
spinless fermions, where the forms of the concerned Hamiltonians remain
unchanged during the bosonization. When bosonizing a spinful 1D fermionic
system, an exotic particle-density-dependent $U(1)$ gauge phase factor emerges from
fermionic statistics among fermions with opposite spin orientations, leading
to one of main challenges in the simulation of the spinful DIII models.

In this paper, we reveal  unambiguously that a 1D HCB lattice can be exploited to faithfully
simulate the proposed DIII model which realizes the nontrivial $\mathbb{Z}_{2}$ topological phase \cite{ZhaoWang-MF}. Most remarkably, with the odd-parity
Kramers doublet ground states being used as the basis of topological qubits,
we demonstrate for the first time that a set of universal quantum gates can in principle
be achieved using these topological bases with the help of this HCB architecture, where we develop a dispersive dynamic
modulation (DDM) approach to mediate the exotic $U(1)$ gauge field
configuration in the bosonized DIII model. Moreover, we propose a
physical implementation scheme based on superconducting quantum circuit
(SQC) \cite{YouNature2011,DevoretSchoelkopf}. Taking the advantages \iffalse advantage \fi  of
flexibility and scalability of SQC \cite{HouckTureciKoch,KochReview}, we
encode HCBs by superconducting transmon qubits and design the nontrivial
coupling through the inductive connection between transmon qubits \cite%
{KochTransmonPRA2007,SchreierPRB2008,MartinisCoupling1,MartinisCoupling2}.
%The feasibility of the DDM method is discussed.
Our estimation also implies that the
effective coupling strengths can be several orders larger than \iffalse that corresponding to the coherence time \fi the decoherence rates of the transmon qubits \cite{BozyigitNatPhys2011,Barends2013},
and therefore the proposed topological operations may be tested with the
current level of technology.

%This paper is organized as follows: We first introduce in  Sec. \ref{Sec HCB} the basic idea of HCB simulation. Then we describe the scheme of implementing universal quantum gates in Sec. \ref{Sec Gates}. In Sec. \ref{Sec SQC} the SQC realization of the proposed schemes is illustrated , and the discussions and conclusion are made in Sec. \ref{Sec Conclusion}.

\section*{Results}
\label{Sec HCB}

\subsection*{The HCB realization of DIII model}

\begin{figure}[tbh]
\begin{center}
    \includegraphics[width=0.48\textwidth]{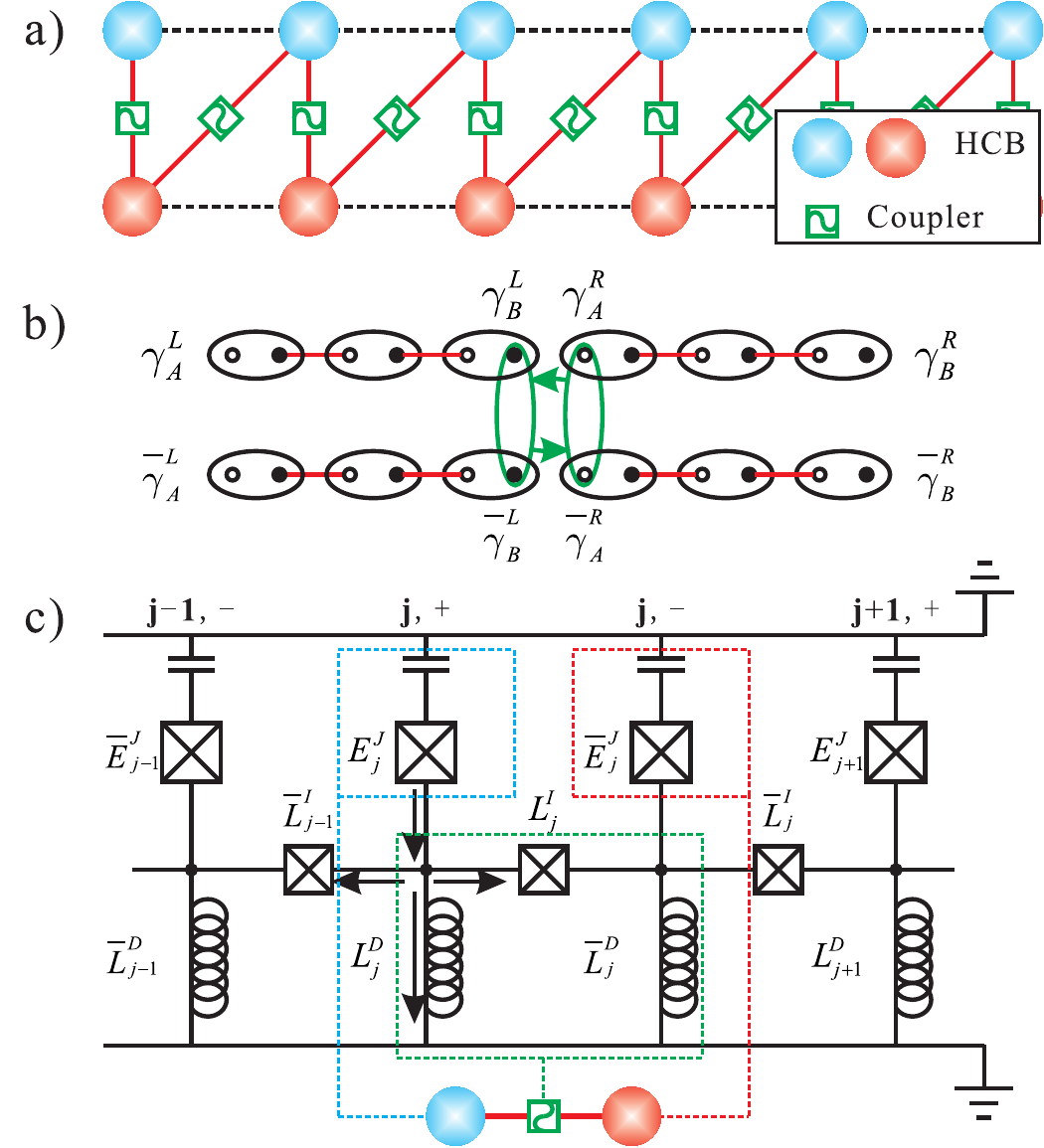}
\end{center}
\caption{\label{Fig Lattice} (a) Schematic view of the 1-D HCB simulator of $H_{\mathrm{DIII}}$.
The solid red line describes the path along which the HCB sites are
physically coupled and the bosonization is performed. Through the DDM of the
nearest neighbor coupling (the green wavelet boxes) the effective
intra-species hopping and pairing (the black dashed lines) can be induced.
The 12-HCB lattice shown is minimal for the demonstration of universal
quantum operations. (b) Realization of nontrivial two-qubit gate in the 12-HCB
lattice. In the situation $w=\Delta $ and $\mu =0$, each HCB site (the ellipse) can be decomposed into two Majorana modes (the filled and the hollow dots) pairing in the bulk. Two topological qubits can be prepared by
\textquotedblleft cutting\textquotedblright\ the whole DIII chain into two
pieces. Through the modulation of $\left( 3,-\right) \longleftrightarrow
\left( 4,+\right) $ coupling, the effective $\protect\gamma %
_{B}^{L}\protect\gamma _{A}^{R}+\bar{\protect\gamma }_{B}^{L}\bar{%
\protect\gamma }_{A}^{R} $ coupling can be induced. (c) Schematic plot of
the array of coupled transmon qubits. }
\end{figure}

An intriguing model Hamiltonian of fermionic 1D topological superconductor in the class DIII
reads \cite{ZhaoWang-MF},
\begin{equation}
H_{\mathrm{DIII}} =\sum_{j\alpha }(-wc_{j,\alpha }^{\dagger }c_{j+1,\alpha
}-i\Delta c_{j+1,\alpha }c_{j,\bar{\alpha }}+\mathrm{h.c.})-\mu ( c_{j,\alpha }^{\dagger }c_{j,\alpha }-1) \label{Eq DIIIModel}
\end{equation}%
where $c_{j,\alpha }^{\dagger }$ is the creation operator of the spin-$
\alpha $ fermion on the $j$th site with $\alpha =\uparrow ,\downarrow $, $w$ is the real-value tight-binding hopping strength, $\mu$ is the chemical potential, $
\Delta $ is the real-value \iffalse real \fi amplitude of p-wave pairing parameter, and $\bar{\alpha}$ denotes the spin component opposite to $\alpha$. Under the basis
transformation $a_{j}=e^{-i\pi /4}(c_{j,\uparrow }+c_{j,\downarrow })/\sqrt{2%
}$ and $\bar{a}_{j}=e^{-i\pi /4}(c_{j,\uparrow }-c_{j,\downarrow })/%
\sqrt{2}$, $H_{\mathrm{DIII}}$ can be rewritten as
\[
H_{\mathrm{DIII}}=H+
\bar{H}
\] with
\begin{small}
\begin{eqnarray}
\begin{array}{c}
H=\sum\limits_{j}(-wa_{j}^{\dagger }a_{j+1}+\Delta a_{j+1}a_{j}+\mathrm{h.c.})-\mu
(a_{j}^{\dagger }a_{j}-1/2)\\
\bar{H}=\sum\limits_{j}(-w\bar{a}_{j}^{\dagger }\bar{a}_{j+1}-\Delta
\bar{a}_{j+1}\bar{a}_{j}+\mathrm{h.c.})-\mu (\bar{a}%
_{j}^{\dagger }\bar{a}_{j}-1/2) \label{F-DIII}
\end{array}
\end{eqnarray}
\end{small}
i.e., $H_{\mathrm{DIII}}$ is decoupled into two Kitaev chains differed by the
signs of their superconducting order parameters (hereafter we denote the
quantities related with the $(+)$-chain as $A$ and their counterparts for the
$(-)$-chain as $\bar{A}$). Although the topology-preserved simulation of a standard Kitaev
model with HCBs has been proposed \cite%
{ImamogluMajorana,WangMajorana,MaoMajorana}, to simulate $H_{\mathrm{DIII}}$
with HCBs is much more challenging than the separate simulations of $H^{+}$ and $H^{-}$ by two
independent HCB chains, because such straightforward scenario cannot result in
the fermionic anti-commutation relation of inter-species. For this, here we exploit a quasi-1D HCB chain shown in Fig.\ref{Fig Lattice}(a),
where the upper/lower array plays the role of the $+/-$ chain,
respectively. The DIII model (\ref{F-DIII}) can be bosonized as

\begin{equation}
\begin{array}{ll}
H&=\sum\limits_{j}(-w\bar{P_{j}}b_{j}^{\dagger }b_{j+1}+\Delta \bar{P_{j}}%
b_{j+1}b_{j}+\mathrm{h.c.})-\mu (b_{j}^{\dagger }b_{j}-\frac{1}{2}) \\
\bar{H}&=\sum\limits_{j}(-wP_{j+1}\bar{b}_{j}^{\dagger }\bar{b}%
_{j+1}-\Delta P_{j+1}\bar{b}_{j+1}\bar{b}_{j}+\mathrm{h.c.}) -\mu (\bar{b}_{j}^{\dagger }\bar{b}_{j}-\frac{1}{2})\label{Eq BosonVersion}
\end{array}
\end{equation}
with the bosonization along the zig-zag path in Fig. \ref{Fig Lattice}(a)
being given by $b_{j}=a_{j}\prod_{s<j}P_{s}\prod_{s<j}\bar{P}_{s}$ and $%
\bar{b}_{j}=\bar{a}_{j}\prod_{s<j+1}P_{s}\prod_{s<j}\bar{P}%
_{s}$. Here $b_{j}$($\bar{b}_{j}$) is the annihilation operators of the
$j$th HCB on the $+$($-$) chain, $P_{s}=\exp (-i\pi a_{s}^{\dag }a_{s})$ [$=\exp (-i\pi b_{s}^{\dag }b_{s})$], and $%
\bar{P}_{s}=\exp (-i\pi \bar{a}_{s}^{\dag }\bar{a}_{s})$. Although this is a nonlocal transformation, the locality of operators of physical noises (or random perturbations) is preserved, and so does the topological stability \cite{Note1}.
%\footnote{This bosonization inherits topological protection against local fluctuations seen from the original fermionic system of Eq.(2), because the locality of the three forms of physical perturbations ($b_{j}^{\dagger}b_{j+1}$, $b_{j+1}b_{j}$, $b_{j}^{\dagger }b_{j}$) in the simulated bosonic system remains unchanged after fermionization. To be concrete in other words, if the transformed terms, which correspond to possible random local (perturbation) terms in the simulated system, are still local ones in the original fermonic picture of topological model after application of the non-local transformation, one can deduce that the simulated system is also topologically protected.}.
It is noticed that the fermionic anti-commutation relation has been converted
to the gauge field factors after bosonization, which in one chain is
given by the particle density of the other chain in a site-wise sense.

To implement the bosonization version of the DIII model (\ref{Eq BosonVersion}), we set that the HCBs on the upper/lower array of Fig.\ref{Fig Lattice}(a) have energy splits $\Omega /\bar{\Omega }$ with $\delta =\Omega -\bar{\Omega }\sim 0.3\Omega $, and the inter-HCB
coupling takes the nearest-neighbor form $H^{\mathrm{L}}=\sum_{j}\mathcal{H}%
_{j}^{\mathrm{L}}+\bar{\mathcal{H}}_{j}^{\mathrm{L}}$ with $\mathcal{H}%
_{j}^{\mathrm{L}}=V_{j}(t)( b_{j}^{\dagger }+b_{j}) (
\bar{b}_{j}^{\dagger }+\bar{b}_{j}) $ and $\bar{%
\mathcal{H}}_{j}^{\mathrm{L}}=\bar{V}_{j}(t)( \bar{b}%
_{j}^{\dagger }+\bar{b}_{j}) ( b_{j+1}^{\dagger
}+b_{j+1}) $ where $\left\vert V_{j}(t)\right\vert ,\left\vert
\bar{V}_{j}(t)\right\vert \simeq 10^{-2} \delta $ are the
coupling strengths which can be modulated harmonically and \textit{in situ}.
We further introduce a dispersive energy threshold $\eta \simeq 10^{-1}\delta $ such that $\left\vert V_{j}(t)\right\vert
,\left\vert \bar{V}_{j}(t)\right\vert \ll \eta \ll \delta $ and
modulate $V_{j}(t)$ and $\bar{V}_{j}(t)$ as
\begin{equation}
\begin{array}{c}
V_{j}(t)=2s_{j}\cos (\eta -\delta )t+2q_{j}\cos (\eta -\epsilon )t \\
\bar{V}_{j}(t)=2\bar{s_{j}}\cos (\eta -\delta )t+2\bar{q_{j}}
\cos (\eta +\epsilon )t%
\end{array}%
\end{equation}%
with $\epsilon =\Omega +\bar{\Omega }$. Based on the parameter choice
\begin{equation}
\begin{array}{cl}
s_{j} =&(-1)^{j-1}s,q_{j}=(-1)^{j-1}q \\
\bar{s}_{j}=&(-1)^{j-1}\bar{s},\bar{q}_{j}=(-1)^{j-1}%
\bar{q},\bar{q}/\bar{s}=-q/s
\end{array}
\end{equation}%
we can directly reproduce the hopping and pairing terms in Eq.(\ref{Eq
BosonVersion}) with $w=-s\bar{s}/\eta $ and $\Delta =s\bar{q}/\eta
$ from the dispersive coupling between $\sum_{j}\mathcal{H}_{j}^{\mathrm{L}}$
and $\sum_{j}\bar{\mathcal{H}}_{j}^{\mathrm{L}}$ (For detailed derivation we refer to \textbf{Methods}).
The obtained amplitudes of $w$ and $\Delta $ can be independently tuned in the range $[10^{-3},10^{-2}]\delta $ by the parameters $(s,q,\bar{s},\bar{q})$. In addition, the chemical potential term in Eq.(\ref{Eq BosonVersion})
can be produced by adding an on-site dispersive radiation $H^{\mathrm{D}%
}=\sum_{j}D_{j}(t)( b_{j}^{\dagger }+b_{j}) +\bar{D}%
_{j}(t)( \bar{b}_{j}^{\dagger }+\bar{b}_{j}) $. A
convenient choice of $D_{j}(t)$ (and similarly $\bar{D}_{j}(t)$) is $%
D_{j}(t)=d_{j}[e^{i\left( \Omega \pm 3\eta \right) }+e^{-i\left( \Omega \pm
3\eta \right) }]$ which results in the a.c. Stark shift $\pm
d_{j}^{2}b_{j}^{\dag }b_{j}/3\eta $. The $3\eta $ setting of the radiation
frequency is to avoid the crosstalk between $H^{\mathrm{D}}$ and $H^{\mathrm{%
L}}$, and the choice of the $3\eta $ sign depends on whether positive or
negative correction of the chemical potential $\mu$ is needed.

\subsection*{MZMs and the universal quantum gates}

The region $|w|>|\mu /2|$
can be identified as the topological phase where the nontrivial MZMs
emerge \cite{Kitaev}. Especially, we focus on the ideal case $w=\Delta $
and $\mu =0$ where the DIII model (\ref{F-DIII}) can simply be reduced to
\begin{equation}
H_{DIII}=\sum_{j}iw\left( \gamma _{j,B}\gamma _{j+1,A}+\bar{\gamma }%
_{j,B}\bar{\gamma }_{j+1,A}\right) ,
\end{equation}%
with Majorana operators $\gamma _{j,A}=i( a_{j}-a_{j}^{\dag }) $,
$\gamma _{j,B}=a_{j}+a_{j}^{\dag }$, $\bar{\gamma} _{j,A}=\bar{a}_{j}+%
\bar{a}_{j}^{\dag }$, and $\bar{\gamma} _{j,B}=-i( \bar{a}_{j}-%
\bar{a}_{j}^{\dag }) $. The Hamiltonian is fully diagonalized
through the pairing of $\gamma _{j,B}$ and $\gamma _{j+1,A}$ (and
simultaneously $\bar{\gamma }_{j,B}$ and $\bar{\gamma }_{j+1,A}$)
in the bulk, leaving four MZMs $\gamma _{A}=\gamma _{1,A}$, $\bar{%
\gamma }_{A}=\bar{\gamma }_{1,A}$, $\gamma _{B}=\gamma _{N,B}$, and $%
\bar{\gamma }_{B}=\bar{\gamma }_{N,B}$ unpaired at the two ends (N being the length of the DIII chain).
The ground states are thus four-fold degenerate with a basis $\{|1\rangle |%
\bar{1}\rangle ,|0\rangle |\bar{1}\rangle ,|1\rangle |\bar{0}%
\rangle ,|0\rangle |\bar{0}\rangle \}$ satisfying $-i\gamma _{A}\gamma
_{B}|0,1\rangle =\pm |0,1\rangle $ and $-i\bar{\gamma }_{A}\bar{%
\gamma }_{B}|\bar{0},\bar{1}\rangle =\pm |\bar{0},\bar{1}%
\rangle $. Here we choose the ground subspace spanned by $|\bullet \rangle
=|0\rangle |\bar{1}\rangle $ and $|\times \rangle =|1\rangle |\bar{%
0}\rangle $ \ as a topological qubit. Such subspace has the odd parity
distinguishing it from the other ground states. For comparison, we recall
that the ground state of the Kitaev model with only one Majorana zero
mode at each end have the opposite (even) fermionic parity~\cite{Kitaev}.

It is shown that a single Kitaev chain can be minimally simulated by a 3--HCB array \cite{WangMajorana}. Therefore, for an easy experimental setup but without loss of generality,  we consider the 12--HCB lattice sketched in Fig.\ref{Fig Lattice}(b) to demonstrate the universal quantum operations in the topological bases. We notice that two DIII chains can be constructed by ``cutting'' the proposed HCB array into two pieces (i.e. by tuning $\bar{s}_{3}=\bar{q}_{3}=0$ and $q_{4}=0$~\cite{Note2}.
%\footnote{Note that turning on or off these coupling parameters does not have any impact on the total ground state energy as MZMs have no contributions to the energy, namely, the total energy is conserved under these operations in the ground state, with the TRS being also preserved.}).
These two subchains, labeled by L and R, have consequently 8 emerged MZMs at the ends (4 for each subchain, see Fig.\ref{Fig Lattice}(b)). For each subchain, the previously established formalism of defining topological qubits can still be exploited. For universal single qubit operation, we take the left subchain as an example by considering $H_{\mathrm{S}}=\mathbf{B}\cdot \hat{\mathbf{S}}^{L}$ with $\mathbf{B}$ as an effective
magnetic field, which is very weak compared with the bulk gap, and $\hat{\mathbf{S}}^{L}%
=\sum_{j=1}^{3}( a_{j}^{\dag },\bar{a}_{j}^{\dag }) \hat{\mathbf{\sigma }}\left( a_{j},\bar{a}_{j}\right) ^{T}$. It has been shown that $\hat{\mathbf{S}}^{L}$ can be regarded as an effective single spin operator acting on the left topological qubit, with $|\bullet \rangle^{L} $ and $|\times\rangle^{L} $ being the two eigenstates of $\mathbf{S}_{z}^{L}$ \cite{ZhaoWang-MF}. Thus through the control of $\mathbf{B}$ the universal single-qubit operation can be achieved. Notably, since the qubits are constructed by the ground states of the whole system, only the global change of the direction of $\mathbf{B}$ may manipulate them, while random local variations are actually irrelevant \cite{Note3}.
%~\footnote{Note that the TRS is weakly broken during a generic adiabatic operation of the single qubit gate, but the operation, similar to the case of the Kitaev chain, is still topologically protected by the particle-hole symmetry of the superconductor, such that this type of gate is no doubt a symmetry-protected topological one unless the superconducting gap vanishes.}. %which implies that this operation is  stable in a topological sense.
The $\mathbf{S}_{x}^{L}$ rotation can be implemented by the resonant dynamic modulation approach: for each $(j,+)\longleftrightarrow (j,-)$ link, we add a resonant tone $V_{j}^{\mathrm{RX}}(t)=2B_{x}(t)\cos \delta t$ to $V_{j}(t)$ with $B_{x}(t)$ the slow-varying (adiabatic)
pulse amplitude satisfying $B_{x}(t_{i})=B_{x}(t_{f})=0$ to induce the resonant inter-species hopping $\sum_{j=1}^{3}B_{x}(t)(b_{j}^{\dag }\bar{b}_{j}+b_{j}\bar{b}_{j}^{\dag })$ which is equivalent to $H_{\mathrm{SX}}=B_{x}(t)\mathbf{S}_{x}^{L}$ in the fermionic picture. The rotation $\exp
(-i\theta \mathbf{S}_{x}^{L})$ with $\theta =\int_{t_{i}}^{t_{f}}B_{x}dt$ can be realized
after $t_{f}$. The $\mathbf{S}_{y}^{L}$ rotation can be similarly realized by changing
the initial phase of the resonant dynamic modulation as $V_{j}^{\mathrm{RY}}(t)=2B_{y}(t)\cos (\delta t-\pi /2)$.
%In addition, the $\mathbf{S}_{z}^{L}$ rotation can be implemented by the previous a.c. Stark shift method of implementing chemical potential terms.

To implement the nontrivial inter-qubit quantum
gate, we modulate $\bar{V}_{3}(t)$ as $\bar{V}_{3}(t)=2\bar{h}%
(t)\cos (\eta -\delta )t$ to induce the effective $\left( 3,+\right)
\longleftrightarrow \left( 4,+\right) $ and $\left( 3,-\right)
\longleftrightarrow \left( 4,-\right) $ hopping which can be written as
\begin{equation}
H_{M}^{\mathrm{T}}=-i\frac{s_{M}}{2}(\gamma _{B}^{L}\gamma _{A}^{R}+%
\bar{\gamma }_{B}^{L}\bar{\gamma }_{A}^{R}),
\end{equation}%
with $s_{M}=s\bar{h}(t)/\eta $. The coupling strength $s_{M}$ can be
controlled by the slow-varying envelope $\bar{h}(t)$ such that $%
\left\vert s_{M}\right\vert \ll \Delta $, $s_{M}(0)=s_{M}(t_{f})=0,$ and $%
\int_{t_{i}}^{t_{f}}s_{M}dt=\pi $. Consequently, we get an unitary
transformation $U=-\gamma _{B}^{L}\gamma _{A}^{R}\bar{\gamma }_{B}^{L}%
\bar{\gamma }_{A}^{R}$ transforming the four basis states as%
\begin{eqnarray}
\begin{array}{c}
|\bullet \rangle ^{L}|\bullet \rangle ^{R} \rightarrow |\times \rangle
^{L}|\times \rangle ^{R},\quad|\times \rangle ^{L}|\bullet \rangle
^{R}\rightarrow -|\bullet \rangle ^{L}|\times \rangle ^{R}, \\
|\bullet \rangle ^{L}|\times \rangle ^{R} \rightarrow -|\times \rangle
^{L}|\bullet \rangle ^{R},\quad|\times \rangle ^{L}|\times \rangle
^{R}\rightarrow |\bullet \rangle ^{L}|\bullet \rangle ^{R},  \notag
\end{array}
\end{eqnarray}%
i.e., $U$ acts as a nontrivial two-qubit gate $-\mathbf{S}_{y}^{L}\otimes \mathbf{S}_{y}^{R}$
which can be exploited to achieve a set of universal quantum gates in
combination with the single-qubit gates mentioned previously.

\subsection*{Physical implementation with superconducting circuits}

We now elaborate in detail
how to implement the present theoretical scheme by SQCs. For the site $%
(j,+)$ we consider a superconducting transmon qubit consisting of a
superconducting quantum interference device (SQUID) with effective Josephson
energy $E_{j}^{\mathrm{J}}$ shunted by a large capacitance $C_{j}$ \cite%
{KochTransmonPRA2007,SchreierPRB2008}, as shown in Fig.\ref{Fig Lattice}(c). The
two characteristic energy scales of the transmon qubit are its anharmonicity
$E_{j}^{\mathrm{C}}=e^{2}/2C_{j}$ and its lowest level splitting $\Omega
_{j}\simeq \sqrt{8E_{j}^{\mathrm{C}}E_{j}^{\mathrm{J}}}$. Here we choose $j$%
-independent $\Omega _{j}/2\pi =\Omega /2\pi =10\,\mathrm{GHz}$ and $E_{j}^{%
\mathrm{C}}/2\pi =E^C/2\pi \simeq 0.75\,\mathrm{GHz}$. The HCB sites on the `$-$' chain
can be similarly constructed with $\bar{\Omega }_{j}/2\pi =\bar{%
\Omega }/2\pi =7.5\,\mathrm{GHz}$ and $\bar{E}_{j}^{\mathrm{C}}\simeq
E_{j}^{\mathrm{C}}$. Due to the large anharmonicity, the presence of the
higher levels of the transmons can only slightly modify the effective
parameters derived below and the transmons can be consequently modeled by
the two-level HCB Hamiltonian $H^{\mathrm{HCB}}=\sum_{j}\Omega b_{j}^{\dag
}b_{j}+\bar{\Omega }\bar{b}_{j}^{\dag }\bar{b}_{j}$. For the
links between neighboring qubits, we exploit the current dividing mechanism
recently studied in experiments \cite{MartinisCoupling1,MartinisCoupling2}.
The transmon qubit design is slightly modified by introducing a small
grounding inductance $L_{j}^{\mathrm{D}}\simeq L_{j}^{\mathrm{%
J}}/2$ with $L_{j}^{\mathrm{J}}=\Phi _{0}^{2}/4\pi ^{2}E_{j}^{\mathrm{J}}$ the
effective inductance of the transmon SQUID at $(j,+)$. A low voltage node is
thus created for each transmon qubit (Fig.\ref{Fig Lattice}(c)), and the
neighboring nodes are connected by coupling SQUIDs with tunable Josephson
inductances $L_{j}^{\mathrm{I}},\bar{L}_{j}^{\mathrm{I}}\simeq L_{j}^{%
\mathrm{J}}/4$. Moreover, the capacitances of the coupling SQUIDs is chosen
to be much smaller than $C_{j}$.

The inter-qubit coupling can be established by the current dividing
mechanism. An excitation current from the $(j,+)$ qubit which can be written
as $I_{j}\simeq \sqrt{\Omega /2L_{j}^{\mathrm{J}}}(b_{j}^{\dagger }+b_{j})$
will mostly flow through $L_{j}^{\mathrm{D}}$ to the ground, with a small
fractions $I_{j+,j-}$ and $I_{j+,(j-1)-}$ flowing to the neighboring qubits $%
(j-1,-)$ and $(j,-)$ through the two coupling SQUIDs. The current $I_{j+,j-}$
in turn generates a flux $\Phi _{j+,j-}=\bar{L}_{j}^{\mathrm{D}%
}I_{j+,j-}$ in the $(j,-)$ qubit. The interaction between the two
qubits can thus be written as
\begin{equation}
{H}_{j}^{\mathrm{L}}=V_{j}(b_{j}^{\dagger }+b_{j})(\bar{b}%
_{j}^{\dagger }+\bar{b}_{j}),
\end{equation}%
where the coupling constant $V_{j}$ can be estimated as $V_{j}\simeq
-(\Omega \bar{\Omega }/L_{j}^{\mathrm{J}}\bar{L}_{j}^{\mathrm{J}%
})^{1/2}\bar{L}_{j}^{\mathrm{D}}L_{j}^{\mathrm{D}}/L_{j}^{\mathrm{I}}$.
Therefore, the a.c. modulation of the penetrating flux bias in the coupling
SQUID loop results in the oscillation of $L_{j}^{\mathrm{I}}$ and in turn
the oscillation of the inter-transmon coupling. Here we should notice that
the modulation frequencies of the coupling SQUIDs should not be higher than
their plasma frequencies, otherwise complicated quasiparticle excitations
would occur. As being mentioned previously, the maximal modulating
frequency of the DDM approach is of the order $\epsilon =\Omega +\bar{%
\Omega }$. Therefore, with parameters being chosen before, this plasma
frequency requirement is guaranteed by the small capacitance of the coupling
SQUIDs.

The d.c. bias of the coupling SQUIDs leads to a static nearest neighbor
qubit-qubit couplings which is estimated to be on the level of $2\pi \lbrack
75,100]\,\mathrm{MHz}$. Since such nearest-neighbor static coupling
strength is much smaller than the energy difference between neighboring
qubits (on the level of $2\pi \times 2.5\,\mathrm{GHz),}$ its influence is
the corrections of the hopping and pairing parameters derived from the DDM
method. The couplings beyond the nearest neighbors should also be estimated:
we notice that the divided current $I_{j+,j-}$ from $(j,+)$ qubit can be
further divided in the $(j,-)$ node to flow through the $(j+1,+)$ node, thus
the next-nearest-neighbor (NNN) $(j,+)\longleftrightarrow (j+1,+)$ static
coupling is induced. Meanwhile, due to the current dividing mechanism at the
low-voltage nodes, the non-nearest-neighbor couplings decay exponentially
with respect to the site distance. The NNN coupling is estimated to be of
the order $2\pi  [15,20]\,\mathrm{MHz}$. To suppress its effect we can
use the two-sublattice stragegy for each of the two HCB legs: the
eigenfrequencies of the four qubits shown in Fig.\ref{Fig Lattice}(c) can
be modified to be $2\pi (7.5,10,7,10.5)\,\mathrm{GHz}$,
respectively. Such modification dose not influence the performance of the
DDM method as we merely need to adjust the modulating frequencies of $%
V_{j}(t)$ and $\bar{V}_{j}(t)$ accordingly. However, the effect of the
d.c. NNN coupling is significantly suppressed by the $0.5\, \mathrm{GHz}$ energy
difference between the NNN transmon qubits.

Based on the static bias, we can add the a.c. modulation pluses on the
coupling SQUIDs with\ $s,q,\bar{s},\bar{q}$ at the order of $2\pi
[10,15]\,\mathrm{MHz}$ and choose the dispersive active region as $%
\eta /2\pi =75\,\mathrm{MHz}$. The resulting dispersive tunneling is
estimated to be on the level $w/2\pi ,\Delta /2\pi \in [1,5]\,\mathrm{MHz}$ which is three orders larger than the reported decoherence
rates of the transmon qubits (in the range $2\pi \lbrack 1,10]\,\mathrm{kHz}
$ \cite{BozyigitNatPhys2011,Barends2013}). Such strong tunnel/pairing allow
us to set the slow varying envelopes on the level $B_{x}(t)/2\pi
,B_{y}(t)/2\pi ,s_{M}/2\pi \in [0.05,0.5]\,\mathrm{MHz}$. Following
this setting, the envelopes are all much smaller than the realized bulk gap such that the high energy
excitations can be omitted, but stronger enough than the decoherence rates
such that the proposed topological quantum operations can be simulated.

\section*{Discussion}

\subsection*{The finite anharmonicity of the transmon qubits}

Here we should notice that the proposed superconducting transmon qubits are not ideal HCBs but anharmonic oscillators. The assumption of regarding them as HCBs with infinite anharmonicity thus needs further investigation.  We recall that the anharmonicity of a transmon qubit is defined by its $\omega_{21}-\omega_{10}$ \cite{KochTransmonPRA2007} and equals its charging energy $E^C$, which is at the level of $2\pi \left[ 0.1, 0.8\right]\,\mathrm{GHz}$. The requirement $E^J/E^C \gg 1$ should be fulfilled because  the qubit becomes more and more sensitive to the background $1/f$ charge noise with decreasing $E^J/E^C$, leading to severe dephasing of the physical qubit \cite{KochTransmonPRA2007}. Here we set $E^C/2\pi=0.75\,\mathrm{GHz}$ such that $E^J/ E^C > 10$ (previously reported high coherence transmon qubits often work in the region $E^J/E^C \approx 20$ \cite{BozyigitNatPhys2011,MartinisCoupling1}). Being larger than the dispersive threshold $\eta/2\pi=75\,\mathrm{MHz}$ by one order of magnitude, such anharmonicity choice prevents the effective excitation of the higher levels of the transmons.

To evaluate the consequence of the finite anharmonictiy, we incorporate the higher levels of the transmon qubits into the ideal HCB model by enlarging the Hilbert space of each HCB to three dimension, i. e. from $\mathbf{span}\left\{ \left|0 \right\rangle_{j,\bar{j}}, \left|1\right\rangle_{j,\bar{j}} \right\}$ to $\mathbf{span}\left\{ \left|0 \right\rangle_{j,\bar{j}}, \left|1\right\rangle_{j,\bar{j}}, \left|2\right\rangle_{j,\bar{j}} \right\}$. Following the same essentials of the DDM method, we come to the result that the inclusion of the anharmonicity leads to parasitic terms of the proposed ideal Hamiltonian, taking the form
\begin{align}
H_{para}&=\sum\limits_{j}(-w_{para}\bar{Q_{j}}b_{j}^{\dagger }b_{j+1} \notag+\Delta_{para} \bar{Q_{j}}
b_{j+1}b_{j}+\mathrm{h.c.}), \\
  \bar{H}_{para}&=\sum\limits_{j}(-\bar{w}_{para}Q_{j+1}\bar{b}_{j}^{\dagger }\bar{b}
_{j+1} -\bar{\Delta}_{para} Q_{j+1}\bar{b}_{j+1}\bar{b}_{j}+\mathrm{h.c.}),
\end{align}
where the strengths of these parasitic terms are of the order
\begin{align}
  |w_{para}|,|\bar{w}_{para}| &\simeq s\bar{s}/E^C \ll w=s\bar{s}/\eta, \\
|\Delta_{para}|,|\bar{\Delta}_{para}| &\simeq s\bar{q}/E^C \ll \Delta=s\bar{q}/\eta,
\end{align}
and
\begin{align}
  Q_j=\exp{\left( -i\pi c_j^{\dagger}c_j\right)},\bar{Q}_j=\exp{\left(-i\pi \bar{c}_j^{\dagger}\bar{c}_j\right)},
\end{align}
with $c_{j}=\left|1\right\rangle_{j}\left\langle 2 \right|$ and $c_j^{\dagger}=\left|2\right\rangle_j\left\langle 1 \right|$. Moreover, as the excitation to the states $|2\rangle_{j,\bar{j}}$ is effectively not excited, we can reduce the obtained parasitic terms back to the original 2D HCB Hilbert space and get
\begin{align}
H_{para}&=\sum\limits_{j}[w_{para}(\frac{1}{2}+\frac{1}{2}\bar{P}_j)b_{j}^{\dagger }b_{j+1} -\Delta_{para} (\frac{1}{2}+\frac{1}{2}\bar{P}_j)b_{j+1}b_{j}+\mathrm{h.c.}], \\
  \bar{H}_{para}&=\sum\limits_{j}[\bar{w}_{para}(\frac{1}{2}+\frac{1}{2}{P}_{j+1})\bar{b}_{j}^{\dagger }\bar{b}
_{j+1}  +\bar{\Delta}_{para} (\frac{1}{2}+\frac{1}{2}{P}_{j+1})\bar{b}_{j+1}\bar{b}_{j}+\mathrm{h.c.}].
\end{align}
We then observe that the parasitic terms contain two types. The first type containing the U(1) gauge factors slightly shift the proposed ideal hopping and pairing constants. This term can be compensated by the refined choice of the pumping parameters. The second type that not containing the U(1) gauge factors can be regarded as local perturbation terms. As estimated before, both the two terms have strength smaller than the proposed band gap by one order of magnitude. Therefore, these parasitic terms results in only the slight correction of the ideal scheme. They cannot destroy the performance of the proposed scheme because they are too small to change the band topology of the system by closing and reopening the gap.

\subsection*{Robustness against imperfection factors}

In realistic experiments, the imperfection accompanies inescapably with the proposed ideal scheme. The deviation of the circuit parameters from their ideal values can be attributed to the errors happened in the fabrication process and the low-frequency noises in the proposed circuit. For the fabrication errors, they cause the deviation of the lowest level splitting $\Omega_j=\sqrt{8E_j^\mathrm{C} E_j^\mathrm{J}}$ of the transmons and the current dividing ratios between neighboring transmons from their proposed values. In recent experiments the fabrication-induced disorder in SQC lattice systems has been significantly suppressed by the developing technology \cite{UnderwoodDisorderPRA2012}. If the fabrication errors are not too large such that the resulted level splits and current dividing ratios are still around their ideal values with small disorder, they can be corrected by simple refinement of the proposed DDM scheme: The deviation of the level splitting can be compensated by the modification of the frequency of the DDM pulses, and the errors of the current dividing ratio can be compensated by the renormalization of the amplitude of the DDM pulses \cite{WangYPChiral2015,WangYPNPJQI2015,WangYPLieb2016}.

The low-frequency $1/f$ noises can also induce fluctuations of the circuit parameters \cite{FlickerRMP2014}. Due to their low frequency property, we can treat the $1/f$ noises as quasi-static, i. e. the noises do not vary during a experimental run, but vary between different runs. As the proposed HCB chain considered consists of only transmon qubits and coupling SQUIDs with very small charging energies, it is insensitive to the charge type $1/f$ noise. Such insensitivity has already been extensively investigated in Ref. \cite{KochTransmonPRA2007}. For the flux $1/f$ noise penetrated in the loops of the SQUIDs, we notice that previous experiments have shown that the flux noise $\delta \Phi$ in a SQUID loop does not vary greatly with the loop size, inductor value, or temperature, and its strength falls in the range $\delta \Phi/\Phi_0 \in [10^{-6}, 10^{-5}]$ \cite{WellstoodFluxNoiseAPL1987,FluxqubitFlickerPRL2006,MartinisFlickerPRL2007,FluxQubit1fGeometryPRB2009}. The consequent fluctuation effects can then be estimated as
\begin{align}
\delta \Omega_{j}/2\pi < 10^{-1}\,\mathrm{MHz}, \quad
\delta s_j, \delta q_j < 10^{-3} \mathrm{MHz},
\end{align}
both of which induce local perturbation to the ideal Hamiltonian with amplitudes much smaller than the proposed band gap of the DIII chain. Such small fluctuations cannot destroy the MZMs as they are too small to change the band topology by closing and reopening the gap (notice that this argument applies to the other parameter fluctuations, and such robustness roots from the topological nature of our scheme). In addition, experiments have shown that the influence of the critical current noise in a large Josephson junction is even smaller than those of the flux $1/f$ noise and can then be safely neglected \cite{MartinisFlickerPRL2007}. Summarizing the analysis above, we then come to the conclusion that our scheme can survive in the presence of the $1/f$ noises in SQC devices.

\section*{Conclusion}
\label{Sec Conclusion}

In conclusion, we have designed a topolgy-preseved HCB simulator for the TRS-invariant DIII topological superconducting chains and proposed a set of universal quantum gates with topological bases through the developed DDM technique. Physical implementation of our proposal with SQCs has also been explored. The present results may pave the way for realizing universal quantum computation with topological stability.

\section*{Methods}
\label{APP DDM}

\subsection*{The derivation of the DDM pulses}

Here we detail the derivation of DDM pulses and the consequent effective Hamiltonian. We exploit the dispersive coupling mechanism which states that, for a system governed by a fast oscillating Hamiltonian $Ae^{i\omega t}+Be^{-i\omega t}$ with $\left\Vert A\right\Vert ,\left\Vert B\right\Vert \ll \omega $, its evolution can be described by the effective Hamiltonian $H^{\mathrm{Eff}}=[A,B]/\omega $ \cite{GoldmanEffPRX2014,GoldmanEffPRA2015} (here the norm $\left\Vert O\right\Vert $ of an operator $O$ is defined as the square root of the largest eigenvalue of $O^{\dag }O$). Our basic idea is that, if we are able to use the $A=b_{j}^{\dagger }\bar{b}_{j}$ term in $\mathcal{H}_{j}^{\mathrm{L}}$ as $A$ and the $\bar{b}%
_{j}^{\dag }b_{j+1}^{\dag }$ and $\bar{b}_{j}^{\dag }b_{j+1}$ terms in $%
\bar{\mathcal{H}}_{j}^{\mathrm{L}}$ as $B$, the resulting commutators $%
[A,B]$ provide the required $b_{j}^{\dagger }b_{j+1}^{\dagger }\bar{P}%
_{j}$ and $b_{j}^{\dagger }b_{j+1}\bar{P}_{j}$ terms. For this purpose
we adopt the rotating frame for which $\mathcal{H}_{j}^{\mathrm{L}}$ and $%
\bar{\mathcal{H}}_{j}^{\mathrm{L}}$ take the forms%
\begin{align}
 \mathcal{H}_{j}^{\mathrm{L}}&=V_{j}( b_{j}^{\dagger }\bar{b}%
_{j}^{\dag }e^{i\epsilon t}+b_{j}^{\dagger }\bar{b}_{j}e^{i\delta
t}+b_{j}\bar{b}_{j}^{\dag }e^{-i\delta t}+b_{j}\bar{b}%
_{j}e^{-i\epsilon t}) ,  \label{Eq j+j-Rotation} \\
\bar{\mathcal{H}}_{j}^{\mathrm{L}}&=\bar{V}_{j}( \bar{b}
_{j}^{\dagger }b_{j+1}^{\dag }e^{i\epsilon t}+\bar{b}_{j}^{\dagger
}b_{j+1}e^{-i\delta t}+\bar{b}_{j}b_{j+1}^{\dag }e^{i\delta t}+\bar{b}_{j}b_{j+1}e^{-i\epsilon t}) .
\end{align}
As being noticed already, the four terms in $\mathcal{H}_{j}^{\mathrm{L}}$
and $\bar{\mathcal{H}}_{j}^{\mathrm{L}}$\ oscillate with frequencies $%
\epsilon ,\delta ,-\delta ~\mathrm{and}~-\epsilon $, respectively.
Meanwhile, the dynamic modulation of $V_{j}(t)$ and $\bar{V}_{j}(t)$
can synthesize the dispersive coupling by shifting the frequencies of the
terms in Eq. (\ref{Eq j+j-Rotation}) upward and downward. For the modulation
$V_{j}(t)=2s_{j}\cos (\eta -\delta )t+2q_{j}\cos (\eta -\epsilon )t$, we
find that $\mathcal{H}_{j}^{\mathrm{P}}=s_{j}b_{j}^{\dagger }\bar{b}%
_{j}+q_{j}b_{j}^{\dagger }\bar{b}_{j}^{\dagger }$ is positively
activated (i.e. it oscillates with frequency $\eta $) and its Hermitian
conjugate $\mathcal{H}_{j}^{\mathrm{N}}=s_{j}b_{j}\bar{b}_{j}^{\dagger
}+q_{j}\bar{b}_{j}b_{j}$ is negatively activated. The other terms in $%
\mathcal{H}_{j}^{\mathrm{L}}$ are de-activated because their frequencies are
far away from the dispersive active region $[-\eta ,\eta ]$ at least by the
order of $\delta $. Also, for the modulation $\bar{V}_{j}(t)=2\bar{%
s_{j}}\cos (\eta -\delta )t+2\bar{q_{j}}\cos (\eta +\epsilon )t$, we
can positively activate $\bar{\mathcal{H}}_{j}^{\mathrm{P}}=\bar{s}%
_{j}\bar{b}_{j}b_{j+1}^{\dag }+\bar{q}_{j}\bar{b}_{j}b_{j+1}$
and negatively activate $\bar{\mathcal{H}}_{j}^{\mathrm{N}}=\bar{s}%
_{j}\bar{b}_{j}^{\dagger }b_{j+1}+\bar{q}_{j}\bar{b}%
_{j}^{\dagger }b_{j+1}^{\dagger }$. The dispersive coupling between $%
\mathcal{H}_{j}^{\mathrm{L}}$ and $\bar{\mathcal{H}}_{j}^{\mathrm{L}}$
results in two effects. The first is the self part
\begin{equation*}
([\mathcal{H}_{j}^{\mathrm{P}},\mathcal{H}_{j}^{\mathrm{N}}]+[\bar{%
\mathcal{H}}_{j}^{\mathrm{P}},\bar{\mathcal{H}}_{j}^{\mathrm{N}}])/\eta
\end{equation*}%
which contains the self-energy corrections of the sites $(j,+)$, $(j,-)$,
and $(j+1,+)$. The more nontrivial terms emerge from the mutual dispersive
coupling
\begin{equation*}
([\mathcal{H}_{j}^{\mathrm{P}},\bar{\mathcal{H}}_{j}^{\mathrm{N}}]+[%
\bar{\mathcal{H}}_{j}^{\mathrm{P}},\mathcal{H}_{j}^{\mathrm{N}}])/\eta
\end{equation*}%
which can be simplified as%
\begin{equation}
\mathcal{H}_{j}^{\mathrm{Eff}}=\frac{s_{j}\bar{s}_{j}}{\eta }%
b_{j}^{\dagger }b_{j+1}\bar{P}_{j}+\frac{s_{j}\bar{q}_{j}}{\eta }%
b_{j}^{\dagger }b_{j+1}^{\dagger }\bar{P}_{j}+\mathrm{h.c}.
\label{Eq J+}
\end{equation}%
Similarly, the dispersive coupling between the $(j,-)\longleftrightarrow
(j+1,+)$ and $(j+1,+)\longleftrightarrow (j+1,-)$ links provides the
effective $(j,-)\longleftrightarrow (j+1,-)$ coupling
\begin{equation}
\bar{\mathcal{H}}_{j}^{\mathrm{Eff}}=-\frac{\bar{s}_{j}s_{j+1}}{%
\eta }\bar{b}_{j}^{\dag }\bar{b}_{j+1}P_{j+1}-\frac{\bar{s}%
_{j}q_{j+1}}{\eta }\bar{b}_{j}^{\dag }\bar{b}_{j+1}^{\dag }P_{j+1}+%
\mathrm{h.c.}.  \label{Eq J-}
\end{equation}%
Notice that all the effective hopping and pairing constants in Eqs.(\ref{Eq
J+}) and (\ref{Eq J-}) can be independently controlled by the pumping
parameters $(s_{j},q_{j})$ and $(\bar{s}_{j},\bar{q}_{j})$.
Through the setting%
\begin{align}
s_{j}& =(-1)^{j-1}s,q_{j}=(-1)^{j-1}q, \\
\bar{s_{j}}& =(-1)^{j-1}\bar{s},\bar{q_{j}}=(-1)^{j-1}%
\bar{q},\bar{q}/\bar{s}=-q/s,  \notag
\end{align}%
the population-dependent phase terms in the bosonic version of $H_{\mathrm{%
DIII}}$ can be directly reproduced from Eqs. (\ref{Eq J+}) and (\ref{Eq J-}).

\section*{Acknowledgments}
We thank T. Mao and D. Zhang for helpful discussions.

\section*{Competing Interests}
The authors declare that they have no competing financial interests.

\section*{Contributions}
Z.D.W proposed the idea. Y.H and Z.Y.X carried out all calculations under the guidance of Z.D.W. Y.X.Z participated in the discussions and the interpretation of the work. Y.H, Y.X.Z, and Z.D.W contributed to the writing of the manuscript.

\section*{Funding}

This work was supported in part by the National Science Foundation of China (Grant No. 11374117), the NKRDP of China (Grant No. 2016YFA0301802), the GRF (HKU173051/14P \& HKU173055/15P), the CRF (HKU8/11G) of Hong Kong, the fellowship of HongKong Scholars Program (Grant No. 2012-80), and the National Fundamental Research Program of China (Grant No. 2011CB922104, No. 2012CB922103, and No. 2013CB921804).

%\bibliography{TP}

\clearpage
\section*{Figure Legends}

\end{document}